\documentclass[notitlepage]{revtex4-1}
\usepackage{graphicx}
\usepackage{amsthm}
\usepackage{amsmath}
\usepackage{amssymb}

\newtheorem{theorem}{Theorem}
\newtheorem{definition}{Definition}
\newtheorem{lemma}{Lemma}

\begin{document}


\title{Exotic Smoothness in Four Dimensions and Euclidean Quantum Gravity}

\date{\today}

\author{Christopher L Duston}
\affiliation{Physics Department, The Florida State University}

\begin{abstract}
In this paper we calculate the effect of the inclusion of exotic smooth structures on typical observables in Euclidean quantum gravity. We do this in the semiclassical regime for several gravitational free-field actions and find that the results are similar, independent of the particular action that is chosen. These are the first results of their kind in dimension four, which we extend to include one-loop contributions as well. We find these topological features can have physically significant results without the need for additional exotic physics.
\end{abstract}

\keywords{General Relativity, Quantum Gravity, Semiclassical Approximations, Exotic Smoothness}

\maketitle

\section{Introduction}
\label{intro}

A key development in quantum field theory was Feynman's path integral approach \cite{Feynman}. It is a generally covariant method to determine quantum amplitudes, and laid the foundations for our understanding of relativistic quantum theory. It is unfortunate that the method that has been so successful in the formulation of quantum field theories (electroweak and QCD) has never been well understood in the gravitational regime. There are several important reasons for this, but many of them are related to the difficulty in dealing with integrals over topological structures \cite{Schleich,Hawking_book}. Such functional integrals must be taken over physically distinct topological structures as well as over gauge-inequivalent field configurations. Some of these issues can be relieved by restricting the path integral to finite sums over extreme points of the action (the semiclassical approach). Of course, each term in the sum must still represent a physically different spacetime, but determining if a spacetime model is different than any other at the level of differentiable structures is highly non-trivial. In addition, it is not known if changing the differentiable structure will actually change observable physics, although this has been strongly conjectured \cite{AMB,AMB-cosmology,Brans, Asselmeyer}. It is this issue that our study focuses on; does the inclusion of inequivalent smooth structures produce different physical results? 


The question of fundamental differences in the structure of manifolds belongs in the realm of topology, and in the past 50 years there have been many discoveries that show the issue is more finely detailed than our intuition suggests. There are several ways that two spaces can be determined to be mathematically ``different'', but not all of these are physical relevant. For instance, general covariance tells us that the laws of physics should not change in different coordinate systems, but what if it is not possible to find differentiable transition functions between locally smoothable coordinate patches? Then there will be explicit differences in the physical results from different regions of such a space. In fact, this results has already been shown to be true for an exampe in 7 dimensions \cite{Schleich} This may have profound implications for our view of gravity and quantum field theory, and is a purely topological notion. These ideas fall into the topic of ``exotic smoothness'', whose inclusion into classical and quantum gravity has been reviewed by \cite{AMB}. The greatest challenge is to find techniques that work in the physically important dimension 4.

The primary goal of this paper is to determine the geometry of a set of 4 dimensional exotic spaces in enough detail to be able to perform a physical calculation; this essentially means determining a metric on them. Once this has been done, because the semiclassical formulation involves a discrete sum rather than an integral, quantities that were previously difficult to calculate become tractable. This will allow us to determine if the addition of exotic structure will have any effect on a calculation. This does not represent a complete determination of the path integral, but it will show explicitly that the inclusion of exotic structures can have an influence on a real physical observable.

We will begin in \S\ref{motevation} by discussing the path integral approach in the semiclassical case and how to formulate the partition function. We will also discuss exotic smoothness and some interesting physical effects that can arise from it. In \S\ref{existence} we present the spacetime models we will be working with and discuss their existence. In \S\ref{volume} we determine the metric and discuss the calculation of volume on these spaces. In \S\ref{section_volume} and \S\ref{section_correlation} we will use several different action functionals to formulate the semiclassical expectation value of volume and the correlation function. In \S\ref{examples} we present explicit examples of our exotic manifolds and discuss the results of the volume and correlation calculations on these spaces. We will finish in \S\ref{1loop} by discussing the one-loop renormalized results, which will illustrate some of the conformal scale dependence of the approach and complete our analysis.

\section{Physical Motivation}
\label{motevation}

The essence of general relativity is Einstein's brilliant insight that gravitation is a manifestation of the geometry of spacetime. Mathematically, this requires a manifold endowed with a metric, the components of the metric being the basic field variables. Our definition of a topological manifold will be taken to be an $n$-dimensional topological Hausdorff space with a countable base that is locally homeomorphic to a subset of $\mathbb{R}^n$ \cite{Straumann}. Except where it is explicitly stated, we will be working with $n=4$. There is an additional structure on a manifold that is relevant for physics and that is to require that it be differentiable. This means that locally the manifold is not just homeomorphic to $\mathbb{R}^n$, but is diffeomorphic to it. This allows us to do calculus in a local region, which is the basis of any reasonable field theory. However, it is the addition of this differentiable (or \textit{smoothness}) structure that is the key point for our discussion.

Frequently, it is tacitly assumed in the physics literature that the existence of a smooth structure is not only guaranteed but that this structure is unique. However, \cite{Milnor1956} discovered the first example of a topological manifold with several \textit{different} smooth structures, which he constructed out of $\mathbb{S}^3$ bundles on $\mathbb{S}^4$. Since then, there have been several examples in high dimension (greater then 4), but the techniques developed (namely the h-cobordism theorem) are not applicable to dimensions less than 5. For dimensions $n<4$, one can show via handlebody decompositions that smooth structures are unique. Paradoxically therefore, the dimension that is most relevant to physics ($n=4)$ has proven to be the most difficult one to study mathematically. It was only later that significantly advanced techniques were available to begin studying exotic structure in dimension 4, and many examples of exotic structures were found \cite{Donaldson}. In fact, it has been shown that there are infinitely many exotic structures on 4 dimensional flat space \cite{Gompf}. Since current results in cosmology strongly point to our universe being flat \cite{WMAP}, it is extremely important that we understand the effect these exotic structures can have on any physical observables. However, it has not yet been possible to put a metric on any of these exotic $\mathbb{R}^4$, only to detect their existence \cite{AMB,Brans,Asselmeyer}. The spaces analyzed in this paper are so far the only 4-dimensional exotic spaces where a metric can be explicitly presented. However, it has recently been shown that knot surgery can be used in 4 dimensions to show some similar results to our own \cite{Asselmeyer_knot}.

At this point, we formally define what we mean by exotic structures.

\begin{definition}
If two manifolds $M,N$ are homeomorphic but non-diffeomorphic, they are \textbf{exotic} to each other.
\end{definition}
The possible implications for physics are immediately obvious; since we rely on calculus to formulate our physical theories, if a single manifold has several inequivalent smooth structures, there may be several inequivalent physical results that can be calculated from the same theory. Not only does this appear to violate the basic principles of relativity, but if we want to perform an experiment to verify a calculation, how could we possibly determine which smooth structure to use? We will discuss several specific applications below, but it should be stressed that despite the name, there is no exotic \textit{physics} here. What we are discussing is exotic \textit{topology} only; the theory is not getting more complicated, we are only recognizing that it has \textit{always} been more complicated than we thought.

\subsection{Path integral formulation of quantum gravity}
The inclusion of exotic smooth structures in quantum gravity occurs at the level of the path integral. Since we are dealing with different structures that may exist over a given spacetime, one anticipates each structure to act as a quantum state of the metric. The standard path integral approach was first developed by \cite{Feynman}, and was extraordinarily successful (with a lot of hard work) at turning quantum mechanics into a relativistically invariant field theory. This further lead to the development of the Standard Model of particle physics, which has passed nearly every experimental test devised. Unfortunately, when directly applied to quantum gravity, this approach leads to a nonrenormalizable theory \cite{Hamber,PS,Srednicki}.

The basic path integral in quantum field theory is the integral over all field configurations and is represented as a probability amplitude, with weights equal to the classical action as the phase of each configuration. In quantum gravity, there are some modifications that need to be made to ensure that the path integral is well-defined. A Wick rotation $t=-\rm i\tau$ is used to take the Lorentzian signature to a spacetime metric with a Riemannian signature, which ensures the amplitude vanishes at infinity. With these considerations, the path integral is

\begin{equation}\label{full_partition}
Z=\int[d g_{ab}]e^{-\frac{1}{\hbar}I[g_{ab}]}.
\end{equation}
In the above expression $I[g_{ab}]$ is the Euclidean classical action corresponding to a specific theory of gravity with no matter content and is a functional of the metric $g_{ab}$. We will consider three possible actions in \S\ref{section_volume}; the Einstein-Hilbert action, the Weyl action, and the spectral action from noncommutative geometry.

Once an action has been specified, the difficulty becomes how to formally integrate over the metrics $g_{ab}$ without double-counting or permitting metrics that are physically unreasonable (for more detail on issues relevant to this construction, see \cite{Hamber} and \cite{Schleich}). If one knew how to do this it would amount to a complete theory of quantum gravity, which does not yet exist. However, in order to use the path integral approach in the current setting we can restrict ourselves to the metrics that will minimize the action and thus contribute the most. This is referred to as the \textit{semiclassical} approach. This general approach has been used with great success to calculate definite results in Euclidean quantum gravity \cite{GibbonsHawking,Hawking}.

If this can be done, the path integral is now a finite sum, and looks formally like the partition function from statistical mechanics:

\begin{equation}
\label{part}
Z=\sum_{i}e^{-\frac{1}{\hbar}I[g^i]},
\end{equation}
where we have suppressed the indices on the metric $g^i_{ab}$ for clarity. In the standard approach, the metrics in this sum are all physically distinct (nondiffeomorphic) and satisfy the same boundary conditions. For instance, one can include a flat solution and a Schwarzschild solution to study black hole thermodynamics \cite{Hawking_book}. In our case, we want to isolate the effect of different smooth structures, so part of our boundary conditions will be to specify a topology. This implies our spaces are all homeomorphic, and it becomes clear how we are to construct the above partition function. Once a single member $g^i$ of the above sum is identified, we want to study the effects of adding all other metrics in the same homeomorphism class but which are nondiffeomorphic (\textit{exotic}). We will also consider one-loop corrections to the partition function in \S\ref{1loop}; these correspond to weighting factors in the above sum.

In some sense, we are factoring the partition function into products whose dynamics do not affect each other. A similar process is done to isolate certain sectors of the standard model, \textit{ie} considering only the electroweak interaction of neutrinos. While we have assumed a given topology for our solutions, we do not consider different \textit{geometries} (metrics) over each smooth structure. For each smooth structure we will be choosing a single metric, and so these metrics will specify the smooth structure as well. Not much is known about how many metrics can be specified on a given smooth structure (outside of conformal transformations, which we will discuss), but it is clear that there is some coupling between smoothness and geometry. For example, \cite{LeBrun} gives an example of a topological 4-manifold with two smoothness structures, one which allows an Einstein metric and one which does not. Therefore in at least some cases, choosing a smooth structure restricts the types of metrics that can be defined on the geometric structure. For an interesting approach in 4 dimensions where no coupling between the metric and the smooth structures is assumed, see \cite{Asselmeyer_knot}.

\section{Exotic Covers of $\mathbb{C}P^2$}
\label{existence}

As previously discussed, if two smooth structures of a given manifold are exotic there exists a homeomorphic but not a diffeomorphic mapping between them. Naturally, directly showing that such a map cannot exist is quite difficult, so the usual approach is to construct smooth invariants of manifolds. When these have different values on homeomorphic manifolds one knows that the respective smooth structures are inequivalent. Unlike topological invariants, such as homotopy and homology groups, smooth invariants come from more sophisticated structures on manifolds such as gauge theory (Yang-Mills, Seiberg-Witten, \textit{etc.}). At the topological level, the main result we will be using is Freedman's classification theorem: 

\begin{theorem}[Freedman's Classification Theorem]
Two smooth, compact, simply-connected 4-manifolds are homeomorphic if and only if they are both spin or both non-spin and their Euler characteristics and signatures are equal.
\end{theorem}

In order to detect differences in smooth structure we one can use Donaldson polynomials, which are constructed out of maps between the 2nd homology groups of the base manifold and the moduli space $M_k$ of irreducible, anti-self-dual, gauge-equivalent connections (see \cite{AMB,Donaldson} for details on this construction). When specializing these polynomial invariants to minimal, simply-connected algebraic surfaces (of which our example will be), we arrive at the key result which we will use to determine the smooth structures in our study.   

\begin{theorem}
\label{div_theorem}
Let $S$, $S'$ be minimal, simply-connected smooth algebraic surfaces of general type with the geometric genus $p_g(S)=p_g(S')\equiv 0$ (mod 2). Then if $f:S\rightarrow S'$ is an orientation-preserving diffeomorphism, the divisibilities of $K_S$ and $K_{S'}$ are equal in integral cohomology.
\end{theorem}

Details of this theorem can be found in \cite{Friedman}. The situation important to our work will be the contra-positive; if the divisibilities of the canonical classes of two algebraic surfaces are different, a diffeomorphism cannot exist. 

We wish to use these invariant properties of manifolds to find families of 4-manifolds that are both exotic and have well-defined metrics. The spaces used in the present study were originally presented by \cite{BK}, who built upon the work of \cite{Salvetti}. The surfaces are iterated covers of the projective plane $\mathbb{C}P^2$.  Each algebraic surface $Y_r$ has two sets of $r$ positive integers, $d_1$,...,$d_r$ and $m_1$,...,$m_r$. They are constructed by starting from the projective plane and repeatedly passing to $r$ coverings each of degree $d_i$ and branched along smooth curves of degree $n_i=d_im_i$ in the plane. Following \cite{Salvetti}, they can be described via branched covering maps $\alpha_i:Y_i \to \mathbb{C}P^2$ (with $\alpha_0=\mathbb{I}$) and cyclic covering maps $\beta_i: Y_i\to Y_{i-1}$ of degree $d_i$ ramified over $\alpha^*_{i-1} (C_i)$. Since each $\alpha_i=\beta_i\circ \beta_{i-1}\circ ...\circ \beta_1$, the covering $\alpha_r$ is of degree $d_1d_2...d_r$.

These surfaces clearly satisfy several of the criteria detailed above: they are algebraic, smooth, minimal, and oriented. It is also fairly easy to see that they are simply-connected \cite{Salvetti}. Another nice feature, which makes these surfaces particularly well-suited for our study is that they admit Einstein metrics. This is because, except for some small values of the parameters, the canonical bundle for a given $Y_r$ is ample. A result by \cite{Aubin} and \cite{Yau} tells us this is enough to prove there is a unique (up to scaling) Einstein metric on such a surface. In fact, the full result of \cite{BK} gives us not only exotic pairs, but entire families of exotic manifolds:

\begin{theorem}
\label{existence_theorem}
For every natural number $k$ there are simply connected topological 4-manifolds $Y_r$ which have at least $k$ distinct structures supporting Einstein metrics. For a given $k$, the ratio $c_1^2/\chi$ is dense in the interval [4,6].
\end{theorem}

Here $c_1$ is the first Chern class and $\chi$ is the Euler characteristic. For convenience we will be using a slightly different set of invariants, the canonical class $K_M=c_1^2(M)$ and the signature $\sigma (M)=(1/3)(c_1^2(M)-2\chi (M))$. Along with the divisibility $n$, these invariants for the above surfaces $Y_r$ were calculated by \cite{BK};

\begin{eqnarray}
K_{Y_r}&=&d_1...d_r\left(\sum_{j=1}^r (d_j-1)m_j-3\right)^2=d_1...d_r n^2,\nonumber\\
\sigma (Y_r)&=&-\frac{1}{3}d_1...d_r\left(\sum_{j=1}^r(d_j^2-1)m_j^2-3\right).
\end{eqnarray}
This also allows us to see the conditions under which our surfaces will have ample canonical bundle. An ample canonical bundle satisfies $K_M\cdot K_M>0$ and $K_M \cdot C>0$ for any curve $C$ \cite{Scorpan}, so the bundle will be ample except for some small values of $(d_i,m_i)$. Being ample is equivalent to being of general type, so our surfaces are of general type in accordance with theorem \ref{div_theorem}.

From a physical point of view, since these manifolds are classical solutions to the Euclidean Einstein equations they represent gravitational instations \cite{Gibbons-Pope}. They can be interpreted as pseudoparticles of the gravitational field, and their symmetry properties can be classified by the above topological invariants \cite{Gibbons-Hawking}.

\section{Metric and Curvature Calculations}
\label{volume}

Now that we have identified the family of exotic manifolds we will be using for our physical model, we now need to find the volume of these covers to calculate the expectation value of volume for a set of them. Since the metric of our base space is well known, we can use the covering functions described in \S\ref{existence} to determine the metric and curvature components of our iterated covers. First we must recall some definitions relevant to the study of complex differential manifolds \cite{Moroianu,Wells}.

\begin{definition}
A \textbf{Hermitian metric} on an almost complex manifold (M,J) is a Riemannian metric h such that h(X,Y)=h(JX,JY) $\forall X,Y\in TM$. The \textbf{fundamental 2-form} of a Hermitian metric is defined by $\Omega(X,Y)=h(JX,Y)$.
\end{definition}

In holomorphic coordinates $z_\alpha$ on Hermitian manifold $(M^{2m},h,J)$ the fundamental form is given by
\begin{equation}
\Omega=i \sum^{m}_{\alpha,\beta=1} h_{\alpha\bar{\beta}} d z_{\alpha}\otimes d\bar{z}_{\beta}.
\end{equation}
In $n$ (complex) dimensions, $\alpha,\beta,...\in\{1,...,n\}$. With the fundamental form we can find the determinant $\Delta$ of the matrix ($h_{\alpha\bar{\beta}}$),

\begin{equation}\label{d}
\frac{1}{n!}\Omega^n=\Delta 2^nd x,~~~d x=\frac{i}{2}d z_1\wedge d\bar{z}_1\wedge ...\frac{i}{2}\wedge d z_n\wedge d\bar{z}_n.
\end{equation}
This gives us a convenient formula for the Ricci tensor in local coordinates,

\begin{equation}\label{Ricci}
R_{\alpha\bar{\beta}}=-\frac{\partial^2 \log \Delta}{\partial z_{\alpha}\partial\bar{z}_{\beta}},
\end{equation}

The metric on the base space $\mathbb{C}P^2$ is the \textit{Fubini-Study metric}. In a local coordinate system specified by $U_j=\{[\zeta_0;...;\zeta_n]|\zeta_j\neq 0\}\subset\mathbb{C}P^n$ and a set of holomorphic coordinate charts $\phi_j:U_j\to\mathbb{C}^n$ defined by

\begin{equation}\phi_j([\zeta_0;...;\zeta_n])=\left(\frac{\zeta_0}{\zeta_j},...,\frac{\zeta_{j-1}}{\zeta_j},\frac{\zeta_{j+1}}{\zeta_j},...,\frac{\zeta_n}{\zeta_j}\right)\end{equation}
for $\zeta_j\neq0$, this metric is

\begin{equation}h=\frac{\sum_{\alpha\beta}(1+|z|^2)\delta_{\alpha\bar{\beta}}-\bar{z_{\alpha}}z_{\beta}}{(1+|z|^2)^2}d z_\alpha\otimes d\bar{z}_\beta\end{equation}
where $|z|^2=|z_1|^2+|z_2|^2+...+|z_n|^2$ \cite{Wells}. 

To calculate the Ricci tensor, we need to find the fundamental form of the metric pulled back over some local coordinate chart to $\mathbb{C}^n$. In fact, things are even simpler; by reading off from the metric we see the fundamental form of $\mathbb{C}P^2$ satisfies $d\Omega =0$. 

\begin{definition}
A complex surface is referred to as \textbf{K\"{a}hler} if the fundamental form is closed, $d\Omega =0$.
\end{definition}

Our base space $\mathbb{C}P^2$ is a K\"{a}hler manifold with an Einstein metric defined on it, sometimes referred to as a K\"{a}hler-Einstein manifold. The following lemma will help us to greatly simplify the calculations of the metric\cite{Moroianu}.

\begin{lemma}
Let $\omega$ be a real 2-form of type (1,1) on a complex manifold $M$. Then $\omega$ is closed if and only if every point in $M$ has an open neighborhood $U\subset M$ such that the restriction of $\omega$ to $U$ equals $i\partial\bar{\partial}u$ for some function $u$ on $U$.
\end{lemma}

Since our fundamental form is closed we set $\Omega=\omega|_{U_j}=i (\phi_j)^*\partial \bar{\partial}u$, so on any K\"{a}hler manifold there is a unique local function $u$ (called the \textit{K\"{a}hler potential}) which completely characterizes the Hermitian metric on that manifold. For the case of $\mathbb{C}P^2$, the K\"{a}hler potential is $u(z_1,z_2)=\log (1+|z_1|^2+|z_2|^2)$. It can be easily shown this manifold is Einstein, $R_{\alpha\bar{\beta}}=3h_{\alpha\bar{\beta}}$, with curvature scalar $R=12$.

On the iterated covers, away from any branch loci the surface is diffeomorphic to an open set in $\mathbb{C}P^2$, where we know the metric. Thus we just need to determine what happens to the curvature in the region of one of the curves. On a $p$-fold covering space $Y_1$ of $\mathbb{C}P^2$, the region around the preimage of the curve $C_1$ will be described by coordinates $(z_1^p,z_2)$. Similarly, on a $q$-fold covering space $Y_2$ of $Y_1$, the region around the preimage of $C_2$ can be chosen to be $(z_1,z_2^q)$. Since these curves have only normal crossing singularities \cite{Salvetti}, on an intersection of these curves we can take the coordinates to be $(z_1^p,z_2^q)$. 

For an arbitrary number $r$ of covers, on any of the iterations $Y_i$, away from the branch loci $C_1\bigcup ...\bigcup C_i$ the surface is locally diffeomorphic to $\mathbb{C}P^2$ and has local coordinates $(z_1,z_2)$. Thus on some higher iteration $Y_j$ of degree $d_j$ ($j>i$), we have coordinates $(z_1^{d_j},z_2)$. Near an intersection with some curve $C_k$ over which the surface $Y_k$ of degree $d_k$ ($k>j$) is branched we can choose coordinates $(z_1^{d_j},z_2^{d_k})$.

Thus our ``top'' iteration $Y_r$ consists of regions diffeomorphic to $\mathbb{C}P^2$ with coordinates $(z_1,z_2)$ and regions where coordinates can be chosen to be $(z_1^{d_i},z_2^{d_j})$ where $d_i,d_j\in \{d_1,...,d_r\}$. We will see that some of the results in this paper will be very dependent on how these coordinates are chosen, but others will not.

Let us now consider some specific choice of coordinates ($z_1^p,z_2^q$) on the space $Y_r$. Then the K\"{a}hler potential becomes

\begin{equation}u'(z_1,z_2)=\log (1+(z_1\bar{z}_1)^p+(z_2\bar{z}_2)^q),\end{equation}
and it is easy to see that this gives a determinant

\begin{equation}\Delta_{pq}=\frac{p^2q^2(z_1\bar{z}_1)^{p-1}(z_2\bar{z}_2)^{q-1}}{(1+(z_1\bar{z}_1)^p+(z_2\bar{z}_2)^q)^3}.\end{equation}

Now using the expression (\ref{Ricci}) we can calculate the Ricci tensor;
\begin{equation}R_{1\bar{1}}=3\frac{p^2z_1^{p-1}\bar{z}_1^{p-1}(1+z^q_2\bar{z}^q_2)}{(1+(z_1\bar{z}_1)^p+(z_2\bar{z}_2)^q)^2},\qquad R_{2\bar{1}}=-3\frac{pqz_1^{p}\bar{z}_1^{p-1}z_2^{q-1}\bar{z}^q_2}{(1+(z_1\bar{z}_1)^p+(z_2\bar{z}_2)^q)^2},\end{equation}
\begin{equation}R_{1\bar{2}}=-3\frac{pq z_1^{p-1}\bar{z}_1^p z^q_2\bar{z}_2^{q-1}}{(1+(z_1\bar{z}_1)^p+(z_2\bar{z}_2)^q)^2}, \qquad R_{2\bar{2}}=3\frac{q^2z_2^{q-1}\bar{z}_2^{q-1}(1+z_1^p\bar{z}_1^p)}{(1+(z_1\bar{z}_1)^p+(z_2^q\bar{z}_2)^q)^2}.\end{equation}

Thus, the metric on the covers is an Einstein manifold with $R_{\alpha\bar{\beta}}=3h_{\alpha\bar{\beta}}$. $\mbox{Tr }h =4$, and we find the curvature in a region around the branch locus is the same as on the base space, $R=12$.

We can understand a little better why the metric is still Einstein by considering the Ricci form of the manifold.

\begin{definition}
The \textbf{Ricci form $\rho$} of a K\"{a}hler manifold is defined by
\begin{equation}\rho (X,Y)=R(JX,Y)~\forall X,Y\in TM.\end{equation}
\end{definition}
Then in local coordinates we have

\begin{equation}\label{Ricciform}
\rho=-i\partial\bar{\partial}\log \Delta,
\end{equation}
and the condition for an Einstein metric $R(X,Y)=\lambda g(X,Y)$ becomes

\begin{equation}\rho=\lambda \Omega\end{equation}
for Einstein constant $\lambda$. Calculating the Ricci form for our iterated covers gives

\begin{eqnarray}
\rho&=&-i\partial\bar{\partial}\log\left[\frac{p^2q^2z_1^{p-1}\bar{z}_
1^{p-1}z_2^{q-1}\bar{z}_2^{q-1}}{(1+z_1^p\bar{z}_1^p+z_2^q\bar{z}_2^q)^3}\right]\nonumber\\
&=&-i\partial\bar{\partial}\left[\log (p^2q^2z_1^{p-1}\bar{z}_1^{p-1}z_2^{q-1}\bar{z}_2^{q-1})-3\log (1+z_1^p\bar{z}_1^p+z_2^q\bar{z}_2^q)\right].\\
\end{eqnarray}
But note that the first term is a closed form,
\begin{equation}\partial\bar\partial\log (z_1^{p-1}\bar{z}_1^{p-1}z_2^{q-1}\bar{z}_2^{q-1})=\partial\left[\frac{p-1}{\bar{z}}d\bar{z}_1+\frac{q-1}{\bar{z}_2}d\bar{z}_2\right]=0.\end{equation}
Thus, the Ricci form is proportional to the fundamental form 

\begin{equation}\rho=3i\partial\bar{\partial}u(z_1,z_2)=3\Omega\end{equation}
and the metric on the cover remains Einstein.

\subsection{The volume of an iterated branched covering of $\mathbb{C}P^2$}
Since the scalar curvature for our iterated branched covers remains constant, it is fairly obvious that a calculation of the action for our model will reduce to a calculation of volume. Since the branch loci are curves of codimension 1, they do not contribute to an integral over the entire space. In addition, away from the curves, the spaces are just $d_1...d_r$ copies of $\mathbb{C}P^2$, so the volume would simply be $d_1...d_r V_0$, where $V_0=2\pi^2$ is the volume of $\mathbb{C}P^2$ \cite{Gibbons-Pope}. We will see that the regions around the branch loci do not have any effect on this volume calculation. To show this, it is sufficient to determine what happens to the volume form on charts around the branch loci.

The volume form of an $n$-dimensional (complex) manifold $M$ can be related to the fundamental form $\Omega$ on $M$ by \cite{Wells}

\begin{equation}d V_M=\frac{1}{n!}\Omega ^n.\end{equation}
Specializing to our $pq$-fold covering, considering the intersection $C_i\bigcap C_j$, and using expression (\ref{d}), we find

\begin{eqnarray}
d V_{Y_r}&=&2^2 \Delta _{pq} \frac{i}{2}d z_1\wedge d\bar{z}_1\wedge \frac{i}{2}d z_2\wedge d\bar{z}_2\nonumber\\
&=&-\frac{p^2q^2(z_1\bar{z}_1)^{p-1}(z_2\bar{z}_2)^{q-1}}{(1+(z_1\bar{z}_1)^p+(z_2\bar{z}_2)^q)^3}d z_1\wedge d\bar{z}_1\wedge d z_2\wedge d\bar{z}_2.
\end{eqnarray}
We need to integrate this over $\mathbb{C}^2$, so using standard coordinate transforms $z_1=x_1+i y_1$, $z_2=x_2+i y_2$ we get

\begin{equation}d V_{Y_r}=4\frac{p^2q^2(x_1^2+y_1^2)^{p-1}(x_2^2+y_2^2)^{q-1}}{(1+(x_1^2+y_1^2)^p+(x_2^2+y_2^2)^q)^3}d x_1\wedge d y_1\wedge d x_2 \wedge d y_2.\end{equation}
This can be put into a more suggestive form by using two sets of polar coordinates, with $x_1^2+y_1^2=r_1^2$ and $x_2^2+y_2^2=r_2^2$.

\begin{eqnarray}
\int_{\mathbb{C}^2}d V_{Y_r}&=&4\int \frac{p^2q^2r_1^{2(p-1)}r_2^{2(q-1)}}{(1+r_1^{2p}+r_2^{2q})^3}r_1d r_1d\theta r_2d r_2d \phi\nonumber\\
&=&4(2\pi)^2\int_{r_1=0}^{\infty}\int_{r_2=0}^{\infty}\frac{p^2q^2r_1^{2p-1}r_2^{2q-1}d r_1d r_2}{(1+r_1^{2p}+r_2^{2q})^3}.
\end{eqnarray}

To compute this integral, first set $u=1+r_1^{2p}+r_2^{2q}$:

\begin{equation}\int_{\mathbb{C}^2}d V_{Y_r}=2(2\pi)^2p^2q\int_{r_1=0}^{\infty}\int_{u=1+r_1^{2p}}^{\infty}\frac{d u}{u^3}r_1^{2p-1}d r_1=4\pi^2pq^2\int_{r_1=0}^{\infty}\frac{r_1^{2p-1}d r_1}{(1+r_1^{2p})^2}.\end{equation}
Now set $v=1+r_1^{d_r}$:

\begin{equation}\int_{\mathbb{C}^2}d V_{Y_r}=2\pi^2pq\int_{1}^{\infty}\frac{d v}{v^2}=2\pi^2pq.\end{equation}

In other words, around an intersection of curves $C_i \bigcap C_j$, the volume is the total degree of the covering times the volume of the base space. On an iterated cover $Y_r$ this would be $d_{r-1}d_rV_{r-2}$, and iterating down to $\mathbb{C}P^2$ we would have $d_1...d_r(2 \pi )^2$. Thus, although the branch locus changes the local representation of the metric, the volume of the total space does not change at all.

\section{The Expectation Value of Volume}
\label{section_volume}
Now that we have determined the geometry of our exotic spaces, we can specify the action functional and use the partition function (\ref{part}) to calculate the semi-classical expectation value of volume. We will consider three different action functionals, and we will show that our results have a similar form in all three cases. In each case we assume a mass-free region (with vanishing stress-energy tensor $T_{ab}=0$). We also define the expectation value of volume to take the form

\begin{equation}
<V>=\frac{\sum_i V_i e^{-I_i}}{\sum_i e^{-I_i}},
\end{equation}
where $V_i$ is the volume of the state $i$ with Euclidean action $I_i$.

\subsection{The Einstein-Hilbert action with $\Lambda\neq 0$}\label{E-H}
This is the action most frequently used in classical relativity and leads to Einstein's original field equations. Under the Euclidean rotation $t=-i\tau$ this is

\begin{equation}
I[h_{ab}]=-\frac{c^3}{16\pi G}\int (R-2\Lambda)d\mu (h_{ab}),\end{equation}
with Riemann curvature $R$, cosmological constant $\Lambda$ and integration measure $d\mu (h_{ab})$, which explicitly depends on the Hermitian metric $h_{ab}$. 

Since we are only interested in solutions which minimize the action, we need to solve the classical field equations \cite{Wald}

\begin{equation}R_{ab}-\frac{1}{2}Rh_{ab}+\Lambda h_{ab}=0.\end{equation}
If $\Lambda=0$, these field equations reduce to $R_{ab}=\frac{1}{2}Rh_{ab}$. However, since $R=12$ and our manifolds are Einstein with $R_{ab}=3h_{ab}$, the field equations cannot be solved in this case. This requires us to consider a non-zero cosmological constant.

If $\Lambda\neq 0$ and the manifold is Einstein, the field equations can be rewritten as 

\begin{equation}R_{ab}=\frac{2\Lambda}{n-2}h_{ab},\end{equation}
Where $n=h_{ab}h^{ab}=4$, the dimension of the space we are considering. Thus, $R=4\Lambda$ and we have

\begin{eqnarray}\label{action}
I[h_{ab}]&=-\frac{2c^3\Lambda}{16\pi G}\int d\mu (h_{ab})\nonumber\\
&=-\frac{c^3\Lambda}{8\pi G}V(h_{ab}).
\end{eqnarray}

Now the connection to the expectation value of volume is through a standard technique in statistical mechanics. Taking the derivative of the natural log of the partition function,

\begin{equation}\frac{\partial \ln Z}{\partial \Lambda}=\frac{\sum_i -\frac{1}{\hbar}\frac{\partial I}{\partial \Lambda}e ^{-\frac{1}{\hbar}I[h^i]}}{Z}=\frac{c^3}{8\pi \hbar G}\frac{\sum_i V(h^i)e^{-\frac{1}{\hbar}I[h^i]}}{Z},\end{equation}
where $V(h^i)$ is the volume of a space with metric $h^i$. This is formally the expectation value of volume for a set of surfaces in a given homeomorphism class. The volume of a surface $Y_r$ with metric $h^i$ is $V(h^i)=2\pi^2\mathfrak{D}_i$, where $\mathfrak{D}_i=d_1...d_r$ is the total degree of the cover. Using the reduced Planck length $l_p=(8\pi G\hbar / c^3)^{1/2}$ and $\Lambda=3$, we can write the expectation value of volume as

\begin{equation}\frac{\partial \ln Z}{\partial \Lambda}=\frac{2\pi^2}{l_p^2}\frac{\sum_i \mathfrak{D}_i e^{6\pi^2\mathfrak{D}_i/l_p^2}}{\sum_i e^{6\pi^2 \mathfrak{D}_i/l_p^2}}=\frac{2\pi^2}{l_p^2}<\mathfrak{D}_{EH}>.\end{equation}
In the next section we will compute this expectation value for specific sets of exotic spaces to see what the overall effect will be for the inclusion of nondiffeomorphic manifolds to the volume.

\subsection{The Weyl action}
The Weyl action is constructed by using the square of the Weyl curvature, $C_{abcd}$, which is the transverse traceless part of the Riemann curvature. In $n\geq 3$ dimensions,

\begin{equation}C_{abcd}=R_{abcd}-\frac{2}{(n-1)(n-2)}h_{a[c}h_{d]b}R+\frac{2}{n-2}(h_{b[c}R_{d]a}-h_{a[c}R_{d]b}).\end{equation}
The Euclidean Weyl action with coupling constant $\alpha_W /16\pi$ is

\begin{equation}
I_W[h_{ab}]=\frac{\alpha_W}{16\pi}\int C_{abcd}C^{abcd} d\mu (h_{ab}).\end{equation}
However, it can be shown that Lagrangians containing a linear combination of the scalars $R_{ab}R^{ab}$ and $R^2$ contain the same dynamical information as the square of the Weyl tensor \cite{DeWitt}. Thus, we take our action to be\cite{Fradkin}

\begin{equation}I_W[h_{ab}]=\frac{\alpha_W}{8\pi}\int (R_{ab}R^{ab}-\frac{1}{3}R^2) d\mu (h_{ab}).\end{equation} 

This action defines \textit{conformal gravity}, which is well known to be a renormalizable quantum theory consisting of 4th derivatives of the metric. In addition, recent work suggests that it does not show the presence of ghosts as other higher-derivative theories do \cite{Bender,Mannheim}. The classical field equation governing the dynamics of conformal gravity is the Bach equation \cite{Bach},

\begin{equation}W_{ab}=\frac{1}{32\pi\alpha_W}T_{ab},\end{equation}
where the Bach tensor $W_{ab}$ is the conformal analogue of the Ricci tensor;

\begin{eqnarray}
W_{ab}&=&-\frac{1}{6}h_{ab}\nabla_c\nabla^cR+\frac{2}{3}\nabla_a\nabla_bR+\nabla_c\nabla^cR_{ab}-2\nabla_c\nabla_{(a}R_{b)}^{~~c}+\frac{2}{3}RR_{ab}+\nonumber\\
&&-2R_a^{~c}R_{cb}+\frac{1}{2}h_{ab}R_{cd}R^{cd}-\frac{1}{6}h_{ab}R^2.
\end{eqnarray}

Although this formulation appears far more complicated than in the Einstein-Hilbert case, since our surfaces are Einstein with constant scalar curvature the Bach tensor in $n$ dimensions reduces to

\begin{equation}W_{ab}=\left(\frac{2}{3}\lambda R-2\lambda^2+\frac{n}{2}\lambda^2-\frac{1}{6}R^2\right)h_{ab}.\end{equation}
This is formally similar to the Einstein condition $R_{ab}=\lambda h_{ab}$. In vacuum, this provides us with the scalar condition that must be satisfied to minimize the action,

\begin{equation}-\frac{1}{6}R^2+\frac{2}{3}\lambda R+(\frac{n}{2}-2)\lambda^2=0.\end{equation}

Thus for $n=4$ the action becomes

\begin{equation}I_W[h_{ab}]=\frac{\lambda \alpha_W}{2\pi} \int (\lambda-\frac{1}{3}R)d\mu (h_{ab}).\end{equation}
Using the Einstein condition $R=\lambda n$ to write this in terms of the scalar curvature gives
 
\begin{equation}I_W[h_{ab}]=-\frac{\alpha_W}{96\pi}R^2 V(h_{ab}).\end{equation}

The expectation value of volume is now the derivative with respect to the scalar curvature,
\begin{equation}\frac{\partial \ln Z}{\partial R}=\frac{R\alpha_W}{48\pi\hbar}\frac{\sum_i V_ie^{-\frac{1}{\hbar}I[h_i]}}{Z},\end{equation}
or with $l_p=(8\pi\hbar / \alpha_W)^{1/2}$ and $R=12$,

\begin{equation}\frac{\partial \ln Z}{\partial R}=\frac{4\pi^2}{l_p^2}\frac{\sum_i \mathfrak{D}_ie^{24\pi^2\mathfrak{D}_i/l_p^2}}{\sum_i e^{24\pi^2\mathfrak{D}_i/l_p^2}}=\frac{4\pi^2}{l_p^2}<\mathfrak{D}_W>.\end{equation}


\subsection{Topological terms in the spectral action}
Another set of actions whose effect one could study is those that contain a topological term, such as the Euler characteristic. Such terms are present in the gravitational Lagrangian derived from the spectral action principal in noncommutative geometry \cite{NCG}. In that case, the additional term is

\begin{equation}R^*R^*=\frac{1}{4}\epsilon_{abcd}\epsilon^{efgh}R^{ab}_{~~ef}R^{cd}_{~~gh}.\end{equation}
The integral of this is the Euler characteristic of a manifold $M$ \cite{spectral},

\begin{equation}\chi(M)=\frac{1}{32\pi^2}\int R^*R^*d\mu.\end{equation}

As noted in \cite{DeWitt}, the variation of this term vanishes and so contains no dynamical information. Thus to study it in a semiclassical calculation we must include a term that gives classical results; such a model can be found in \cite{NCG}, where for high energies the spectral action has a topological term that takes the asymptotic form

\begin{equation}I_S[h_{ab}]=\frac{\alpha_S}{16\pi}\int (3C_{abcd}C^{abcd}-\frac{11}{6}R^*R^*)d\mu(h_{ab}).\end{equation}
Using the identity \cite{Fradkin}

\begin{equation}\label{identity}
R_{ab}R^{ab}-\frac{1}{3}R^2=\frac{1}{2}C_{abcd}C^{abcd}-\frac{1}{2}R^*R^*,\end{equation}
we can write the spectral action as

\begin{eqnarray}
I_S[h_{ab}]&=&\frac{3\alpha_S}{8\pi}\int (R_{ab}R^{ab}-\frac{1}{3}R^2+\frac{7}{36}R^*R^*)d\mu(h_{ab})\nonumber\\
&=&\frac{3\alpha_S}{8\pi}\left[\int (R_{ab}R^{ab}-\frac{1}{3}R^2)d\mu(h_{ab})+\frac{7}{36^2\pi^2}\chi (M)\right]\nonumber\\
&=&\frac{3\alpha_S}{8\pi}\left[-\frac{R^2}{12}V(h_{ab})+\frac{7}{36^2\pi^2}\chi (M)\right].
\end{eqnarray}

Now our volume calculation proceeds in a similar manner;

\begin{eqnarray}
\frac{\partial \ln Z}{\partial R}&=&\frac{3\alpha_S R}{4\cdot 16\pi\hbar}\frac{\sum_i V(h^i)e ^{-\frac{1}{\hbar}I_S[h^i]}}{Z}\nonumber\\
&=&\frac{9\pi^2}{l_p^2}\frac{\sum_i \mathfrak{D}_ie^{72\pi^2\mathfrak{D}_i/l_p^2}e^{-21\chi(M_i)/36^2\pi^2l_p^2}}{\sum_i e^{72\pi^2\mathfrak{D}_i/l_p^2}e^{-21\chi(M_i)/36^2\pi^2l_p^2}}
\end{eqnarray}
with $l_p=(8\pi\hbar / \alpha_S)^{1/2}$. However, since each manifold in the exotic set is homeomorphic, they will have the same Euler characteristic and that term will drop from the expectation value. Thus the quantity we will be concerned with is

\begin{equation}
<\mathfrak{D}_S>=\frac{\sum_i \mathfrak{D}_ie^{72\pi^2\mathfrak{D}_i/l_p^2}}{\sum_i e^{72\pi^2\mathfrak{D}_i/l_p^2}}.\end{equation}

Therefore for all three of the actions we are considering, the expectation value of volume will be given by the expectation value of the total degree of the covering,

\begin{equation}<\mathfrak{D}_I>=\frac{\sum_i \mathfrak{D}_ie^{\beta_I\pi^2\mathfrak{D}_i/l_p^2}}{\sum_i e^{\beta_I\pi^2\mathfrak{D}_i/l_p^2}},\end{equation}
where $\beta_I=\{6,24,72\}$, depending on the action. It is rather remarkable that regardless of the choice of action the results of this calculation only differ by the weighting of the respective spaces. Additionally, this shows that quantum effects on these exotic spaces would be more prominent in the Einstein-Hilbert case rather than either the conformal or noncommutative case.

\section{Correlation Functions}
\label{section_correlation}
Upon choosing an action we now have a well-defined partition function (\ref{part}) and can use the usual techniques of quantum field theory to compute interesting quantities such as correlation functions. If we promote our partition function to a generating functional by adding a source term $J^{\alpha\beta}(x)$ we get

\begin{equation}Z[J^{\alpha\beta}]=\sum_i e^{\frac{\alpha}{\hbar}\int (\mathcal{L}(h^i)+J^{\alpha\beta}(x)h^i_{\alpha\beta}(x))d\mu(h^i)},\end{equation}
with $i$ indexing our exotic spaces and a choice of Lagrangian $\mathcal{L}$ with coupling constant $\alpha$. We can define the two-point correlation function as \cite{PS,Srednicki}

\begin{equation}G(h_{\gamma\delta}(x_1),h_{\rho\sigma}(x_2))=\frac{1}{Z[J^{\alpha\beta}=0]}\left(-i\frac{\delta}{\delta J^{\gamma\delta}(x_1)}\right)\left(-i\frac{\delta}{\delta J^{\rho\sigma}(x_2)}\right)Z\Biggl{|}_{J^{\alpha\beta}=0}.\end{equation}
The functional derivative here will be taken to be

\begin{equation}\frac{\delta J^{\alpha\beta}(x_1)}{\delta J^{\gamma\delta}(x_2)}=\delta^\alpha_\gamma \delta^\beta _\delta \delta^4 (x_1-x_2).\end{equation}
The result of this calculation is

\begin{equation}\label{correlation}
G(h_{\gamma\delta}(x_1),h_{\rho\sigma}(x_2))=-\left(\frac{\alpha}{\hbar}\right)^2\frac{\sum_i h^i_{\rho\sigma}(x_2)h^i_{\gamma\delta}(x_1)e^{\frac{\alpha}{\hbar}\int \mathcal{L}(h^i)d\mu (h^i)}}{\sum_i e^{\frac{\alpha}{\hbar}\int \mathcal{L}(h^i)d\mu (h^i)}}.
\end{equation}

This can be interpreted as the quantum-mechanical amplitude for the geometry to change from being represented by $h_{\gamma\delta}$ at $x_2$ to $h_{\rho\sigma}$ at $x_1$. The matrix element will be the square of this quantity,

\begin{eqnarray}\label{probability}
M(h(x_1),h(x_2))&=&|G(h_{\gamma\delta}(x_1),h_{\rho\sigma}(x_2))|^2\\
&=&\left(\frac{\alpha}{\hbar}\right)^4\frac{n^2\sum_i e^{\frac{2\alpha}{\hbar}\int \mathcal{L}(h^i)d\mu (h^i)}+\Omega(x_1,x_2)}{(\sum_i e^{\frac{\alpha}{\hbar}\int \mathcal{L}(h^i)d\mu (h^i)})^2},
\end{eqnarray}
where the only coordinate-dependent piece is the cross term

\begin{equation}\label{omega}
\Omega(x_1,x_2)\equiv 2\sum_{i<j}h^i_{\rho\sigma}(x_2)h^{j\rho\sigma}(x_2)h^i_{\gamma\delta}(x_1)h^{j\gamma\delta}(x_1)e^{\frac{\alpha}{\hbar}[\int \mathcal{L}(h^i)d\mu (h^i)+\int \mathcal{L}(h^j)d\mu (h^j)]}.
\end{equation}

The major difference with this calculation and the previous expectation value calculations is that this depends on the details of the components of the metric, and therefore the choices of curves $C_i$ over which the cover is branched. The total integral of the Lagrangian appears in the exponential as before, but the correlations will be dependent on the location of the observer relative to the branch loci in each space. Thus there will be several different results depending on what location on the manifold is being discussed; we will explore these calculations in the next section.

\section{Explicit Examples}
\label{examples}
To find examples of exotic pairs in our family of iterated branched covers of $\mathbb{C}P^2$ and calculate their contribution to the expectation value of volume, a computer search was employed. We know from the result of theorem \ref{existence_theorem} that such surfaces should be plentiful for certain ratios of invariants. However, for physical reasons we would like to restrict our search to ``simple'' manifolds only. As such we define the \textit{complexity} $\mathfrak{C}$ as follows. For a given surface $Y_r=(d_1,m_1)...(d_r,m_r)$ define $d=max(d_1,...,d_r)$ and $m=max(m_1,...,m_r)$. We therefore define the complexity as

\begin{equation}\mathfrak{C}(r,d,m)=md^r.\end{equation}
Note the relation between this quantity and the volume of a surface. For all surfaces with $r$ iterated covers and maximum covering degree $d$, the maximum possible volume is $2\pi^2 d^r$. Thus the complexity is related to the volume, which is the physical quantity with which we are concerned.

Some of the results of the computer search are given in tables \ref{2fold}, \ref{nfold}, and \ref{3fold}. We sampled all surfaces with complexity $\mathfrak{C}\leq 64$ and grouped them into homeomorphic families. For our complete search we found 1001 families with 2 members and 32 families with more than 2. A sampling of the families with two members is given in table \ref{2fold}. We show in detail one of the simplest examples, the pair $Y_1=(4,3)$ and $Y_4=(2,1)(2,1)(2,1)(2,3)$. The invariants are

\begin{eqnarray*}
K_{Y_1}&=&4\cdot(3\cdot 3-3)^2=144,\\
K_{Y_4}&=&2\cdot 2\cdot 2\cdot 2\cdot (1\cdot 1+1\cdot 1+ 1\cdot 1 +1\cdot 3-3)^2=144.\\
\sigma(Y_1)&=&-\frac{1}{3}4\cdot \left[(4^2-1)\cdot 3^2 -3\right]=-176,\\
\sigma(Y_4)&=&-\frac{1}{3}2 \cdot 2\cdot 2\cdot 2\cdot [(2^2-1)\cdot 1+(2^2-1)\cdot 1+(2^2-1)\cdot 1+\\
&&\qquad +(2^2-1)\cdot 3^2-3]=-176.\\
n(Y_1)&=&3\cdot 3-3=6,\\
n(Y_4)&=&1\cdot 1+1\cdot 1+ 1\cdot 1 +1\cdot 3-3=3.\\
\mathfrak{C}(1,4,3)&=&12,\\
\mathfrak{C}(4,2,3)&=&48.\\
\end{eqnarray*}

For these two surfaces our total degrees are $\mathfrak{D}_1=4$ and $\mathfrak{D}_2=2^4$. For the expectation value of degree, this gives

\begin{equation}<\mathfrak{D}_I>=\frac{4e^{\beta_I\pi^2\cdot 4}+16e^{\beta_I\pi^2\cdot 16}}{e^{\beta_I\pi^2 \cdot 4}+e^{\beta_I\pi^2 \cdot 16}}=16\frac{1+\frac{1}{4}e^{-\beta_I\pi^2\cdot 12}}{1+e^{-\beta_I\pi^2\cdot 12}}\approx 16=\mathfrak{D}_2,\end{equation}
with $l_p=1$. Thus at the Planck scale the expectation value of volume is dominated by the larger volume. We will see in the next section that when we take the one-loop corrections into account this result will be strongly dependent on the conformal scale of the metric.

How this result changes depending on specific volumes can be explored by performing a Taylor expansion and assuming $\sum e ^{\beta_I(\mathfrak{D}_i-\mathfrak{D}_1)}<<1$ where $\mathfrak{D}_1$ is the largest multidegree in the set:

\begin{equation}<\mathfrak{D}_I>\approx \mathfrak{D}_1\left( 1-\sum_{j=2}(1-\frac{\mathfrak{D}_j}{\mathfrak{D}_1})e ^{\beta_I(\mathfrak{D}_j-\mathfrak{D}_1)}\right).\end{equation}
Thus the relevant quantity is the relative difference between the volumes of the exotic spaces and not the ratio. Of course, since the values of $\mathfrak{D}_j$ above are integer-valued, at very large volumes the possibility of measurable contributions would seem to increase. In fact, since \cite{BK} proved their initial result using a tower of large covering degrees, it is likely that this is true. In addition, they showed that there are exotic families with $k$ members for any $k\in \mathbb{N}$. The expectation value of volume for a large-$k$ family might have significant contributions from spaces other then the largest. However, our method of computer search was specifically aimed at finding \textit{simple} surfaces, and at least for $\mathfrak{C}\leq 64$ the contribution is small.

Another feature of theorem \ref{existence_theorem} is that the ratio of invariants $c_1^2/\chi$ is dense in the interval $[4,6]$, which corresponds to $0\geq\sigma\geq -K$. Although not all our values fall in this range (for instance, our simple example above does not), most of them do.

To study the correlation function (\ref{correlation}), consider the specific example of the two spaces $Y_1$ and $Y_4$ given above. Let us consider a region around the intersection of two curves $C_1,C_2\in Y_4$, and choose the origin to be at the intersection, $x_1=0$. Now without specifying anything about $Y_1$ we can see that the coordinate-dependent piece of the matrix element (\ref{omega}) vanishes, since $h^{Y_4}_{\alpha\beta}=0$. Thus the correlation function in this case is

\begin{eqnarray*}
M(h(x_1),h(x_2))&=&16\left(\frac{\alpha}{\hbar}\right)^4\frac{e^{\frac{2\alpha}{\hbar}\int \mathcal{L}(h^{Y_4})d\mu(h^{Y_4})}+e^{\frac{2\alpha}{\hbar}\int \mathcal{L}(h^{Y_1})d\mu(h^{Y_1})}}{(e^{\frac{\alpha}{\hbar}\int \mathcal{L}(h^{Y_4})d\mu (h^{Y_4})}+e^{\frac{\alpha}{\hbar}\int \mathcal{L}(h^{Y_1})d\mu (h^{Y_1})})^2}\\
&=&\frac{16}{l_p^8}\frac{1}{1+\eta}.
\end{eqnarray*}
where we have set
\begin{equation}\eta\equiv\frac{2e ^{l_p^{-2}(\int \mathcal{L}(h^{Y_4})d\mu (h^{Y_4})+\int \mathcal{L}(h^{Y_1})d\mu (h^{Y_1}))}}{e^{2l_p^{-2}\int \mathcal{L}(h^{Y_4})d\mu (h^{Y_4})}+e^{2l_p^{-2}\int \mathcal{L}(h^{Y_1})d\mu (h^{Y_1})}}> 0.\end{equation}

As comparison, consider a more general situation with a set of $k$ exotic spaces, this time assuming we are looking in a local chart on a manifold that is away from the branch locus. It is then locally diffeomorphic to $\mathbb{C}P^2$. In this case the metric products in expression (\ref{omega}) are all contractions of the metric, so we get

\begin{equation}\Omega(x_1,x_2)=2n^2\sum_{i<j}e ^{\frac{\alpha}{\hbar}\int \mathcal{L}(h^i)d\mu (h^i)}e ^{\frac{\alpha}{\hbar}\int \mathcal{L}(h^j)d\mu (h^j)}.\end{equation}
The matrix element in this case is

\begin{eqnarray*}
M(h(x_1),h(x_2))&=&\left(\frac{\alpha}{\hbar}\right)^4\frac{n^2\sum_i e^{\frac{2\alpha}{\hbar}\int \mathcal{L}(h^i)d\mu (h^i)}+2n^2\sum_{i<j}e ^{\frac{\alpha}{\hbar}\int \mathcal{L}(h^i)d\mu (h^i)}e ^{\frac{\alpha}{\hbar}\int \mathcal{L}(h^j)d\mu (h^j)}}{(\sum_i e^{\frac{\alpha}{\hbar}\int \mathcal{L}(h^i)d\mu (h^i)})^2}\\
&=&\frac{16}{l_p^8}.
\end{eqnarray*}

This result is interesting in comparison to the previous one; they are both independent of the distance between $x_1$ and $x_2$, but the probability for the geometry to change in a region away from the branch loci is actually larger than near the intersections of the curves. This presents the counterintuitive result that quantum effects may actually be larger in regions of the manifold away from the singularities. It would be very interesting to study this effect for other sets of exotic spaces to determine whether this is a general feature of exotic manifolds or only for our specific case.

\section{One-loop Renormalization}\label{1loop}
In the above analysis of the semiclassical partition function (\ref{part}) we assumed that the only contribution came from the minima of the action, which in this study was given by a family of exotic structures. However, each of these minina may not contribute equally to the full partition function (\ref{full_partition}) if the shape of the quantum action near each minima is different. In order to obtain a better approximation to the full partition function, we adopt the \textit{stationary phase approximation}. We expand the Euclidean action and the metric about each minima,

\begin{equation}
I[g^i]\backsimeq I[h^{i}]+I_2[\bar{h}^i]+\mathcal{O}(h^3)\nonumber
\end{equation}
\begin{equation}
g^i_{ab}=h^{i}_{ab}+\bar{h}^i_{ab},
\end{equation}
where $\bar{h}^i_{ab}$ is a small perturbation about the $i$th minima $h_{ab}^i$, and $I_2[\bar{h}^i]$ is quadratic in the small perturbation. Following \cite{Hawking_book} and \cite{Gibbons}, the generating functional for each minima can then be expressed as

\begin{equation}
\log Z=-\frac{1}{\hbar}I[h]+\log \int \mathcal{D}h \exp \left(-\frac{1}{\hbar}I_2[\bar{h}]\right),
\end{equation}
where the quadratic term looks like

\begin{equation}
I_2[\bar{h}]=\frac{1}{2}\int \bar{h}^{ab}A_{abcd}\bar{h}^{cd} d\mu (h).
\end{equation}
Here $A_{abcd}$ is a second order differential operator. It has a large number of zero eigenvalues, since the action is invariant under a diffeomorphism

\begin{equation}
x_a \rightarrow x_a+\epsilon\eta_a\nonumber\nonumber
\end{equation}
\begin{equation}
g_{ab}\rightarrow g_{ab}+2\epsilon\nabla_{(a}\eta_{b)}.
\end{equation} 
Thus, following the standard technique of Fadeev-Popov determinants, we add to this another second-order differential operator $B$ as a gauge-fixing term. This operator must satisfy $B_{abcd}(\bar{h}^{ab}+2\epsilon\nabla^{(a}\eta^{b)})=0$, and we can choose the harmonic gauge \cite{Hawking_book,Gibbons}. The operator $A+B$ will not have any zero eigenvalues, but we need to remove the eigenvalues of $B$ that we have added, which we do by dividing by the determinant of an operator $C$ (the ghost fields) which are pure gauge transformations, $\bar{h}_{ab}=2\nabla_{(a}\eta_{b)}$. Now expressing these functional integrals as determinants \cite{Srednicki}, we get

\begin{equation}
\log Z=-\frac{1}{\hbar}I[h]-\frac{1}{2}\log \det \left(\frac{A+B}{2\pi\mu^2\hbar}\right)+\log \det \left(\frac{C}{2\pi\mu^2\hbar}\right).
\end{equation}
Although we have removed the zero eigenvalues, these determinants may still be divergent if the eigenvalues increase without bound. These can be regulated by using the zeta function technique \cite{Gibbons,Hawking_Zeta} if one can ensure that $A+B$ has only a finite number of negative eigenvalues. Then we can form a generalized zeta function from the eigenvalues $\{\lambda_n,L\}$ for an operator $L$ as

\begin{equation}
\zeta(s,L)=\sum_n \lambda^{-s}_n(L).
\end{equation}
A basic property of the zeta function is that it can be analytically extended to a meromorphic function which is regular at $s=0$, and thus we can express the determinant of an operator as a function of the derivative of the zeta function at $s=0$:

\begin{equation}
\det L=\exp(-\zeta'(0,L)).
\end{equation}
We can write our operator $A+B=-F+G$, where $F$ acts only on the trace of $h_{ab}$ and $G$ acts only on the trace-free parts of $h_{ab}$ (see \cite{Hawking_book} for details on this decomposition). Since these act on different subspaces, and the operators each have even rank, we can now write $\det (A+B)=\det(F)\det(G)$. For each of the $p$ negative eigenvalues of $F,G$, and $C$, we will have to rotate the contour of integration to ensure the convergence of the path integral. This will introduce a factor of $i^p$ into our action. Thus, our functional integral will look like

\begin{equation}
\log Z=-\frac{1}{\hbar}I[h]+\frac{1}{2}\left(\zeta'(0,F)+\zeta'(0,G)-2\zeta'(0,C)\right)-\frac{1}{2}\log (2\pi\hbar\mu^2)+\frac{1}{2}ip\pi.
\end{equation}
We can remove the dependence on the renormalization parameter $\mu$ by assuming that it is independent of the scale of the background metric\cite{Gibbons}. Under a conformal transformation $\tilde{h}=\Omega^2h$, we have the following:

\begin{equation}
I[\tilde{h}]=\Omega^2h\nonumber\\
\end{equation}
\begin{equation}
\lambda _n(\tilde{L})=\Omega^{-2}\lambda _n(L)\nonumber
\end{equation}
\begin{equation}
\zeta'(s,\tilde{L})=\Omega^{2s}\log \Omega \zeta(s,L)+\Omega^{2s}\zeta'(s,L)\nonumber
\end{equation}
\begin{equation}
\log \tilde{Z}=\log Z+(1-\Omega^2)\frac{1}{\hbar}I[h]+\frac{1}{2}\gamma\log\Omega.
\end{equation}
Here we have defined

\begin{equation}
\gamma:=\zeta(0,F)+\zeta(0,G)-2\zeta(0,C),
\end{equation}
which carries all the additional one-loop corrections in the stationary phase approximation. Note that for $n$ classical solutions (exotic structures in this study), we can write the functional integral like

\begin{equation}
Z=\sum_i W_1[h^{ i},\Omega]\exp\left(-\frac{1}{\hbar}I[h^i]\right),
\end{equation}
where we now have a weighing factor $W_1=\exp({-\frac{1}{2}\gamma \log\Omega})$ which represents the one-loop corrections. This formulation of renormalization in quantum gravity, where the partition function is given as a sum of classical solutions weighted by their one-loop terms, was first suggested by \cite{Hawking}.

In order to determine the weighting factors $W_1$, we must calculate the parameter $\gamma$. In \cite{Hawking_book,Gibbons} this parameter is calculated using the spectral geometry results of \cite{Gilkey}. For the Einstein-Hilbert action with $\Lambda\neq 0$,

\begin{equation} 
\gamma=\int \left(\frac{53}{720 \pi^2}C_{abcd}C^{abcd}+\frac{763}{540\pi^2}\Lambda^2\right)d\mu(h).
\end{equation}
Using the relation (\ref{identity}) and our previous results from the Einstein-Hilbert case we get

\begin{equation}
\gamma=\frac{73}{60\pi^2}\Lambda^2 V(g_i).
\end{equation}

Now referring to \S\ref{E-H} we can write the partition function as

\begin{equation}
Z=\sum_i \exp \left[ \left(-\frac{1}{2}a\log\Omega \Lambda^2+l_p^{-2}\Lambda\right)V[g_i]\right],
\end{equation}
where $a:=73/60$ and the reduced Plank length $l_p$ as in \S\ref{E-H}. Note the essential difference between this expression and the first; the exponential is now quadratic in the cosmological constant, and so a partial derivative of the partition function with respect to it will not yield the formal expectation value of volume. However, since the exponential is linear in the scale factor $\log\Omega$, we can use that to define the expectation value of volume at the one-loop level to be

\begin{equation}
\langle V \rangle:=\frac{1}{a\Lambda^2}\frac{\partial \ln Z}{\partial \log \Omega}=\frac{V_0}{a\Lambda^2}\frac{\sum_i \mathfrak{D}(g_i)\exp (6\pi^2\bar{\Lambda}\mathfrak{D}(g_i))}{\sum_i \exp (6\pi^2\bar{\Lambda}\mathfrak{D}(g_i))}.
\end{equation}
where we have defined 

\begin{equation}
\bar\Lambda=l_p^{-2}-\frac{3}{2}a\log\Omega.
\end{equation}
In these equations we have used $\Lambda=3$ and $V(h^i)=2\pi^2\mathfrak{D}_i$ as before. 

We can now study the expectation value of volume as a function of the conformal scale factor $\Omega$. It turns out that the basic results from \S\ref{examples} are completely different if the scale factor becomes very large. There are several regimes of the scale factor we can investigate.

\begin{figure}\label{V_fig}
\includegraphics[angle=270,scale=0.5]{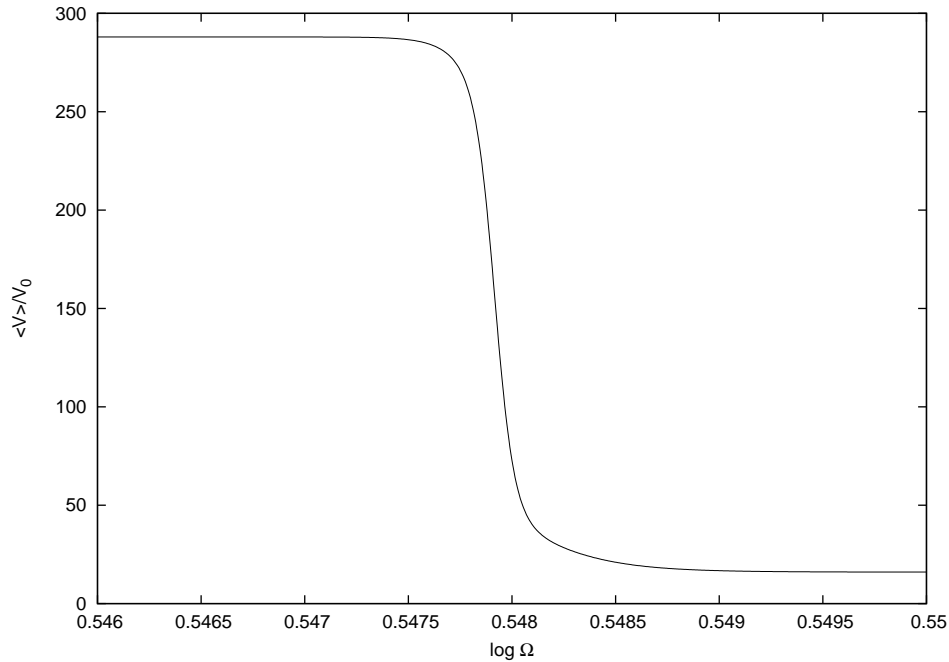}
\caption{The expectation value of volume as a function of conformal scale factor $\Omega$. The dramatic shift in behavior as the scale factor crosses the critical value $\Omega_c=\exp (2 l_p^2 / 3a)$ is clearly seen. Here $V_0=2\pi^2 / a\Lambda^2$, the set of spaces is $\{16,50,72,200,288\}$ and $l_p=1$.}
\end{figure}

\begin{itemize}
\item $\Omega=1$. Then $\log\Omega=0$, $\bar{\Lambda}=l_p^{-2}$ and we return to the original result of \S\ref{examples}.
\item $\Omega<<1$. Then $\bar{\Lambda}>>1$, and the expectation value of volume is even more strongly dominated by the solution with the larger volume.
\item $\Omega=\Omega_c=\exp (2/ 3al_p^2)$. This is the critical case $\bar{\Lambda}=0$. With the weighting factors all equal to one, in this case all the solutions contribute equally to the expectation value. We end up with a simple mean volume, which in the case of an example space $\{16,50,72,200,288\}$ the expectation value of volume is

\begin{equation}
<V>=125.2\frac{2\pi^2}{a\Lambda^2}.
\end{equation}
\item $\Omega>>\Omega_c$. In this case $\bar{\Lambda}<0$ and the exponent in the expectation value of volume is negative. This changes which volumes dominate the expectation value - now the smallest volumes will have the largest effect. A plot of the expectation value of volume which clearly shows these regions of behavior is given in Fig. \ref{V_fig}.
\end{itemize}

\section{Conclusion} 
\label{conclusion}

This paper has shown that one can explicitly compute the contribution to a typical observable due to the inclusion of exotic smooth structures in Euclidean quantum gravity. This represents the first time such a calculation has been done in dimension 4. Although the actual spaces used for this study are somewhat unphysical, it is hoped that some of the techniques illustrated here will lead to further explorations on this topic. 

We found that for semiclassical calculations on iterated branched covers of $\mathbb{C}P^2$, the quantity of interest was a simple expectation value of covering degree, independent of the particulars of the Lagrangian chosen. Since the differences in volume of our exotic spaces were large, such a calculation was dominated by the space with the largest volume, and so the inclusion of exotic structures would have little effect on a practical calculation. However, this result was unique to this set of spaces; a similar model using a different set of exotic structures could have very different results. In addition, the inclusion of the one-loop terms in the functional integral showed that our results were very dependent on the choice of conformal scale for the metric. The next natural step in this study would be to find a set of exotic spaces in 4 dimensions whose physical properties are more similar to each other, and so calculations of expectation values and correlation functions would show a larger effect. Since our exotic spaces were completely specified by their volumes, which varied greatly, they were mostly dominated by a single space. It is not known if this is a general statement about exotic structure, but it is likely unique to this specific set of exotic manifolds. It is our hope that this first step into dimension 4 will lead to a better understanding of the influence of exotic topology in physics and how we can use it to resolve some of the key questions surrounding classical and quantum gravity.  

\begin{acknowledgements}
I would like to thank M. Marcolli for great assistance on this project, as well as E. Klassen, J. Klauder, L. Reina, and O. Vafek for many useful discussions.
\end{acknowledgements}

\bibliography{ref}{}

\pagebreak
\appendix
\section{Sets of Exotic Families}

\begin{table}[h]
\caption{A sample of homeomorphic families of exotic pairs with $\mathfrak{C}\leq 64$.}
\begin{center}
{\small
\begin{tabular}{|c|c|cc|cc|}
\hline
Surface 1&Surface 2&$c_1^2$&$\sigma$&$n_1$&$n_2$\\
\hline
(4,3)&(2,1)(2,1)(2,1)(2,3)&144&-176&6&3\\
(6,3)&(2,1)(2,1)(2,1)(3,3)&864&-624&12&6\\
(10,3)&(2,1)(2,1)(2,1)(5,3)&5760&-2960&24&12\\
(2,6)(2,7)&(2,1)(8,1)&400&-336&10&5\\
(9,15)(10,15)&(10,8)(16,8)&5715360&-1208160&252&189\\
(2,2)(2,3)&(2,1)(2,1)(2,1)(2,1)&16&-48&2&1\\
(2,3)(2,8)&(2,1)(2,1)(2,1)(2,4)&256&-288&8&4\\
(4,4)(4,5)(4,6)&(4,2)(6,2)(6,3)&112896&-24576&42&28\\
(2,7)(3,3)(8,2)&(2,4)(2,5)(3,1)(3,1)(3,3)&27648&-7488&24&16\\
(2,1)(2,4)(2,5)(2,5)(3,3)&(2,1)(2,4)(3,1)(3,2)(3,2)&15552&-4320&18&12\\
\hline
\end{tabular}}
\label{2fold}
\end{center}
\end{table}

\begin{table}
\caption{All homeomorphic families with 4 or 5 members out to complexity $\mathfrak{C}\leq 64$. Each family is complete out to $\mathfrak{C}\leq 128$.}
\begin{center}
{\footnotesize
\begin{tabular}{|c|cc|c|}
\hline
Members of Homeomorphic Family&$c_1^2$&$\sigma$&$n$\\
\hline
(2,3)(4,7)(4,7)&56448&-15936&42\\
(2,13)(3.10)(3.13)&&&56\\
(3,7)(4,4)(6,1)&&&28\\
(2,1)(2,1)(3,1)(4,1)(6,2)&&&14\\
\hline
(2,3)(5,2)(5,4)&28800&-8400&42\\
(2,9)(2,8)(3,9)&&&40\\
(2,1)(2,1)(2,5)(3,3)(3,5)&&&20\\
(2,1)(2,1)(2,1)(5,1)(5,2)&&&12\\
\hline
(2,3)(3,4)(3,4)(5,5)&116640&-26400&36\\
(2,11)(2,12)(2,14)(5,5)&&&54\\
(2,1)(2,3)(5,3)(8,2)&&&27\\
(3,1)(3,4)(5,1)(8,1)&&&18\\
\hline
(2,6)(4,7)(4,7)&64800&-16800&45\\
(2,3)(5,4)(5,5)&&&36\\
(2,8)(3,1)(3,4)(4,5)&&&30\\
(2,7)(4,2)(5,1)(5,1)&&&18\\
(3,1)(4,1)(4,1)(6,2)&&&15\\
\hline
(2,3)(5,7)(5,8)&180000&-45600&60\\
(2,9)(4,10)(4,13)&&&75\\
(3,4)(3,5)(8,5)&&&50\\
(5,1)(5,2)(8,3)&&&30\\
(3,1)(3,2)(4,5)(8,1)&&&25\\
\hline
\end{tabular}}
\label{nfold}
\end{center}
\end{table}

\pagebreak
\begin{table}\small
\caption{A sample of homeomorphic families with more then 2 members with $\mathfrak{C}\leq 64$. Each family is complete out to $\mathfrak{C}\leq 128$.}
\begin{center}
{\small
\begin{tabular}{|c|cc|c|}
\hline
Members of Homeomorphic Family&$c_1^2$&$\sigma$&$n$\\
\hline
(4,12)(8,6)&180000&-47200&75\\
(5,9)(10,3)&      &      &60\\
(4,5)(2,3)(4,2)(5,1)(5,5)&&&30\\
\hline
(4,7)(4,11)&41616&-13584&51\\
(3,1)(3,7)(4,7)&&&34\\
(5,4)(2,1)(2,1)(3,1)(3,2)(4,4)&&&17\\
\hline
(2,3)(2,8)(8,4)&41472&-13056&36\\
(3,7)(4,1)(6,2)&&&24\\
(5,4)(2,1)(2,1)(2,1)(4,2)(4,4)&&&18\\
\hline
(2,6)(3,7)(6,5)&63504&-16464&42\\
(4,1)(7,1)(7,2)&&&18\\
(5,4)(2,1)(2,5)(3,1)(3,5)(4,2)&&&21\\
\hline
(3,4)(3,5)(6,5)&86400&-21600&40\\
(5,1)(5,2)(6,3)&&&24\\
(5,4)(2,1)(3,1)(3,2)(3,5)(4,2)&&&20\\
\hline
(2,7)(2,8)(4,4)&9216&-3072&24\\
(2,4)(2,5)(3,1)(3,4)&&&16\\
(5,4)(2,1)(2,3)(2,3)(2,5)(4,1)&&&12\\
\hline
(2,7)(3,7)(3,7)&18432&-5568&32\\
(2,3)(2,4)(2,5)(4,5)&&&24\\
(5,3)(2,1)(2,4)(2,4)(3,1)(3,4)&&&16\\
\hline
(4,2)(4,2)(4,3)&20736&-5376&18\\
(2,4)(2,5)(3,4)(3,5)&&&24\\
(4,4)(3,1)(3,2)(4,1)(4,2)&&&12\\
\hline
(2,7)(5,3)(8,2)&72000&-16320&30\\
(2,1)(2,4)(3,3)(3,4)(5,1)&&&20\\
(5,5)(2,1)(2,3)(4,1)(4,1)(5,2)&&&15\\
\hline
(2,3)(2,4)(2,4)(5,4)&23040&-6720&24\\
(2,1)(3,3)(3,4)(5,1)&&&16\\
(4,5)(2,1)(4,1)(4,1)(5,2)&&&12\\
\hline
(2,3)(3,5)(3,5)(5,5)&144000&-30720&40\\
(2,1)(4,4)(4,4)(5,2)&&&30\\
(5,5)(2,2)(3,1)(3,1)(4,3)(5,2)&&&20\\
\hline
(2,5)(3,4)(3,5)(5,4)&116640&-23520&36\\
(2,4)(4,2)(4,4)(5,2)&&&27\\
(5,5)(2,1)(3,1)(3,2)(4,2)(5,2)&&&18\\
\hline
(2,3)(4,2)(4,2)(4,3)&56448&-11904&21\\
(2,4)(2,4)(2,5)(3,4)(3,5)&&&28\\
(5,4)(2,2)(3,1)(3,2)(4,1)(4,2)&&&14\\
\hline
(2,16)(16,2)&34848&-12000&33\\
(2,7)(3,6)(3,14)&&&44\\
(2,8)(3,1)(3,6)(4,1)&&&22\\
\hline
\end{tabular}}
\label{3fold}
\end{center}
\end{table}

\end{document}